\documentclass[pre,aps,reprint,footinbib,longbibliography]{revtex4-1}
\usepackage{graphicx}
\usepackage{amssymb}
\usepackage{amsmath}
\usepackage{times}

\begin{document}

\title{Quasi phase reduction 
of all-to-all strongly coupled $\lambda-\omega$ oscillators near incoherent 
states}

\author{Iv\'an Le\'on}
\author{Diego Paz\'o}
\affiliation{Instituto de F\'{\i}sica de Cantabria (IFCA), CSIC-Universidad de 
Cantabria, 39005 Santander, Spain}

\date{\today}

\begin{abstract}
The dynamics of an ensemble of $N$ weakly coupled limit-cycle oscillators can be 
 captured by their $N$ phases using standard phase reduction techniques. 
However, it is a phenomenological fact that all-to-all strongly coupled limit-cycle oscillators 
may behave as ``quasiphase oscillators'', 
evidencing the need of novel reduction strategies.
We introduce here quasi phase reduction (QPR), a scheme suited for identical oscillators with polar symmetry 
($\lambda-\omega$ systems).
By applying QPR we achieve a reduction 
to $N+2$ degrees of freedom: $N$ phase
oscillators interacting through one independent complex variable.
This ``quasi phase model'' is asymptotically valid in the neighborhood 
of incoherent states, irrespective of the coupling strength. The effectiveness 
of QPR is illustrated in a particular case, an ensemble of Stuart-Landau oscillators,
obtaining exact stability boundaries of uniform and nonuniform incoherent states
for a variety of couplings. 
An extension of QPR beyond the neighborhood of incoherence is also explored. Finally, a general QPR
model with $N+2M$ degrees of freedom is obtained for 
coupling through the first $M$ harmonics.
\end{abstract}

  \maketitle

\section{Introduction} 

Dynamical reduction is a concept of paramount importance in nonlinear dynamics \cite{KuraNakao19},
which may be used to reduce the number of degrees of freedom or 
to transform the evolution equations into a canonical form. Classical reduction techniques 
include adiabatic elimination \cite{Hak83}, center-manifold
reduction \cite{Guckenheimer}, and phase reduction \cite{Kur84,nakao16,monga19,pietras19}. 
The latter has been crucial to configure our comprehension of
oscillatory media and coupled self-sustained oscillators. 

Large ensembles of coupled self-sustained oscillators are found in a variety of domains 
ranging from biology and technology to the social sciences, see e.g.~\cite{HI97,Win80,Str03,PikRos15} and references therein.
It is well established that, if the coupling among $N$ limit-cycle oscillators is weak,
then phase reduction can be applied \cite{Kur84} and the dynamics becomes 
reliably described by $N$ phase oscillators.
This approach yields a minimal description of emergent phenomena in all-to-all coupled
oscillators as, for instance, collective synchronization \cite{Win67,Kur75,PRK01},
quasiperiodic partial synchronization (QPS) \cite{Vre96,clusella16} or 
nonuniform incoherent states (NUISs) \cite{leon19}.

If the coupling is strong, however, phase reduction is not applicable as evidenced by
several forms of collective chaos in globally coupled oscillators, 
which  clearly elude phase reduction \cite{HR92,NK93}.
However, there are situations in which the oscillators, despite being strongly coupled,
still resemble phase oscillators, as their ordering on top of a
closed curve is preserved in time. 
Straightforward examples are
states in which the mean field vanishes, such
that each oscillator evolves as if it was uncoupled from the others.
For identical oscillators
these states are called incoherent, or
`phase-balanced configurations' 
if $N$ is finite \cite{DB14}.
The uniform incoherent state (UIS) ---also called `splay state' for finite $N$--- 
is the simplest form of incoherence.
This was encountered long time ago in arrays of Josephson junctions \cite{wiesenfeld89,aronson91},
populations of model neurons \cite{abbott93,treves93,WS96} (with the name of asynchronous state), and other systems.
In contrast, other phase-balanced configurations, i.e.~NUISs, 
have attracted much less attention. We are only aware that coexistence of different NUISs
is nowadays being investigated in the context of some engineering applications \cite{sinha17,sinha18}.
Apart from incoherent states,
there are more complex phenomena such as QPS, modulated 
QPS, or pure collective chaos in which identical
oscillators behave as ``quasiphase oscillators'' on top of an unsteady closed curve \cite{NK93,NK95,CP19}.
Recent advances extending standard phase reduction beyond the first order
do not appear to be practical enough even to cover the moderate coupling regime \cite{leon19,wilson_ermentrout_prl19}. 
Alternative methods based on phase-amplitude reduction or isostables fall short
in the dimensionality reduction actually achieved \cite{castejon13,wilson18,monga19,WilsonErmentrout19}.

In this paper we present quasi phase reduction (QPR), a dynamical reduction method to capture
the 
dynamics of all-to-all coupled identical limit-cycle oscillators near incoherent states.
For standard phase reduction the zeroth order corresponds to tuning the coupling to zero.
In our new approach the incoherent states play the role of zeroth-order solutions, and
the mean field will be the ``small quantity'' of our theory.
Moreover, the number of oscillators is irrelevant, 
it may be either finite or infinite.
The QPR method only covers identical $\lambda-\omega$ oscillators (two-dimensional systems with polar symmetry),
but still, it is conceptually appealing since it yields a significant dimensionality 
reduction from $2N$ to $N+2$ degrees of freedom. The reduced system consists
of $N$ phase oscillators and one complex-valued variable. 
Thereupon we can calculate analytically the stability boundary of incoherent states.
Moreover, we explore an extension of QPR, keeping the $N+2$ degrees of freedom,
which correctly pinpoints a saddle QPS at moderate coupling in 
a specific model. Finally, general QPR with $N+2M$ degrees of freedom is derived for
coupling through the $M$th harmonic.
Throughout this paper the correctness of our approach
is confirmed by numerical simulations with a popular $\lambda-\omega$ system called Stuart-Landau oscillator.

The paper is organized as follows. In Sec.~II we introduce the $\lambda-\omega$ oscillator and the isochrons.
Incoherent states in a particular system of globally coupled $\lambda-\omega$ oscillators 
are reviewed in Sec.~III for illustrative purposes.
Section IV presents QPR for a family of coupling functions. The results in Sec.~IV
are applied to Stuart-Landau oscillators in Sec.~V. Sections \ref{sec_next} and VII extend the results in Sec.~IV
beyond the lowest order, and to other coupling functions, respectively. The conclusions are summarized
in Sec.~VIII.


\section{$\lambda-\omega$ oscillator}

In this work we restrict ourselves to oscillators of the $\lambda-\omega$ type 
\cite{kopell73,Win80}. 
These are 2-dimensional systems with rotational symmetry,
which admit the following 
representation
of the evolution equations in polar coordinates:
	\begin{subequations} \label{lw}
		\begin{eqnarray}
		\dot{r}=\lambda(r)r ,\\
		\dot{\phi}=\omega(r) .
		\end{eqnarray}
	\end{subequations}
The overdot denotes time derivative as usual. Without lack of generality we 
assume
the existence of a stable limit cycle at $r=1$, i.e.~$\lambda(1)=0$. Moreover, 
the natural frequency of the oscillator is $\Omega=\omega(1)$. 
The attraction rate to the limit cycle is given by the second 
Floquet exponent $\Lambda=\left.\frac{d\lambda}{dr} \right|_{r=1}< 0$. 
Alternatively to Eq.~\eqref{lw}, we can 
work with the complex variable $A=re^{i\phi}$, such that the 
$\lambda-\omega$ oscillator obeys:	
	\begin{equation}\label{freeosc}
	\dot{A}=f(A) ,
	\end{equation}
where function $f$ satisfies $f(Ae^{i\alpha})=e^{i\alpha}f(A)$. 
For simplicity, it is convenient to assume that $f$ can be expressed as a 
series of the form
	\begin{equation}\label{fexpan}
	f(A)=\sum_{n=-\infty}^{\infty}f_n |A|^nA ,
	\end{equation}
where $f_n$ are complex coefficients.	
The existence of an attractive limit cycle of frequency $\Omega$ implies
$\sum_n n\operatorname{Re}(f_n)=\Lambda$ and $\sum_n f_n=i\Omega$. 
Common instances of $\lambda-\omega$ systems contain a small number of 
nonzero coefficients $f_n$ in Eq.~\eqref{fexpan}.
If only, $f_0$ and $f_2$ are nonzero (with $\mathrm{Re}(f_0) = - \mathrm{Re}(f_2)>0$) 
we have the well-known Stuart-Landau oscillator \cite{Kur84}, the normal form of a supercritical 
Hopf bifurcation. Adding other nonzero terms we get, for instance, 
the normal form of the generalized (Bautin)
Hopf bifurcation if $f_4\ne0$ \cite{Guckenheimer}, or  
the slow-amplitude dynamics of a parametric feedback oscillator, as used in 
micro- and nano-electromechanics, if $f_{-1}\not=0$ \cite{matheny17}.

\subsubsection{Isochrons}
To account for the effect of perturbations, 
phase reduction approaches require extending the definition of the phase
away from the limit cycle \cite{Win80,Kur84,nakao16,pietras19}. To do so, we seek a phase variable 
$\theta$, such that $\dot\theta=\Omega$ holds 
in the whole basin of attraction, not only on the limit cycle. The 
`isochron' is defined as the set of points that
convergence to the same `asymptotic phase' on the limit cycle.
For $\lambda-\omega$ systems polar symmetry yields
a relation between the phase 
$\theta$ and the polar coordinates of the form \cite{Win80}
\begin{equation}\label{iso}
\theta(r,\phi)=\phi-\chi(r), 
 \end{equation}
with $\chi(1)=0$. The phase dynamics satisfies
$\dot{\theta}=\dot{\phi}-\frac{d\chi}{dr}\dot{r}$, and
imposing $\dot{\theta}=\Omega$, we solve the equation for $\chi(r)$:
	\begin{equation}\label{defiso}
\chi(r)=\int_1^{r}\frac{\omega(\hat{r})-\Omega}{\lambda(\hat{r})\hat{r}}d\hat{r}
=\int_1^{r}\frac{\sum_n \operatorname{Im}(f_n)\hat{r}^{n}-\Omega}{\sum_n 
\operatorname{Re}(f_n) \hat{r}^{n+1}}d\hat{r}
	\end{equation}
Depending on the specific oscillator type considered a closed analytical solution of 
$\chi(r)$
may or not exist. However, if deviations from the limit cycle are small
it is enough to know the first coefficient of the Taylor expansion of $\chi$ 
around $r=1$
	\begin{equation} 
	\chi(1+\delta r)=\chi_0 \delta r+O(\delta r^2) , \nonumber
	\end{equation}
where $\chi_0=\left.\frac{d\chi}{dr}\right|_{r=1}$.
Differentiating Eq.~\eqref{defiso} and 
evaluating the limit $r\to1$ by L'H\^opital's rule,
we get:
	\begin{equation}
		\chi_0=
		\frac{\sum_n n \operatorname{Im}(f_n)}{\sum_n 
(n+1)\operatorname{Re}(f_n)}=\frac{\sum_n n \operatorname{Im}(f_n)}{\Lambda}. \nonumber
	\end{equation}
This expression together with $\sum_n n\operatorname{Re}(f_n)=\Lambda$, obtained before,
can be cast in a compact form:
\begin{equation} \label{nfn}
\sum_{n=-\infty}^\infty n f_n= 
\Lambda(1+i\chi_0) . 
\end{equation}

\subsection{Stuart-Landau oscillators}
\label{SL}

In this paper we asses the validity of our theoretical findings
with the Stuart-Landau oscillator, which
is a universal representation (via center-manifold reduction) 
of systems in the neighborhood of a Hopf bifurcation.
It reads:
\begin{equation}\label{StLa}
\dot{A}=A-(1+i c_2) |A|^2A .
\end{equation} 
The limit cycle at $|A|=1$ has a second Floquet exponent
$\Lambda=-2$. Moreover,
the isochrons are logarithmic spirals
$\theta=\phi-c_2\ln r$,  where $c_2$ in Eq.~\eqref{StLa} is 
the so-called nonisochronicity parameter.
Therefore $\chi_0=c_2$ for this system.

\section{AN EXAMPLE: THE Mean-field complex Ginzburg-Landau equation} \label{secCGLE}

Before presenting the QPR method, it is instructive to recall a well-studied 
model of globally coupled $\lambda-\omega$ oscillators in which incoherent states 
are observed:
The mean-field complex Ginzburg-Landau equation (MF-CGLE). It 
consists of $N$ diffusively coupled Stuart-Landau oscillators \cite{HR92,NK93}:
 \begin{equation}\label{CGLE}
 \dot{A}_j= A_j - ( 1 + i c_2) |A_j|^2 A_j
 +\epsilon(1+ic_1) \left(\overline{A}- A_j\right) ,
 \end{equation}
 here constants $\epsilon$ and $c_1$ determine the strength and the reactivity of the coupling, 
 respectively; and $\overline{A}=\frac{1}{N}\sum_{k=1}^{N}A_k$. 
The MF-CGLE is a discretization of the complex Ginzburg-Landau
equation on a fully connected lattice. The last term of Eq.~\eqref{CGLE} is
a discrete version of the Laplacian on such a lattice.
The MF-CGLE is a prototype of system with many degrees of freedom and a rich repertoire
of collective behaviors. In addition to full synchrony, UIS, NUISs, QPS, and clustering, the system 
displays several forms of chaos and has attracted considerable attention over 
the years \cite{HR92,NK93,NK94,NK95,banaji99,TGC09,KGO15,CP19}.

For better comparison with the QPR theory,
 it is convenient 
 to absorb the local term $\epsilon (1+ic_1) A_j$.
 Specifically, setting 
 \begin{equation}
 \kappa=\frac{\epsilon}{1-\epsilon} , \nonumber
 \end{equation} 
rescaling time ($t\rightarrow \frac{t}{1-\epsilon}$), and 
 going to a rotating frame with rescaled amplitude
 ($A_j\rightarrow 
 \frac{A_j}{\sqrt{1-\epsilon}}e^{-i (\epsilon c_1 +c_2)t}$) we get:
 \begin{equation}\label{MFCGLE}
 \dot{A_j}=(1+i c_2) (1-|A_j|^2)A_j+\kappa (1+i c_1) \overline{A}
 \end{equation} 
 At variance with the Stuart-Landau Eq.~\eqref{StLa}, the linear coefficient has nonzero imaginary part, 
 as we have adopted a rotating frame such that $\Omega=0$.

\begin{figure}
		\includegraphics[width=\linewidth]{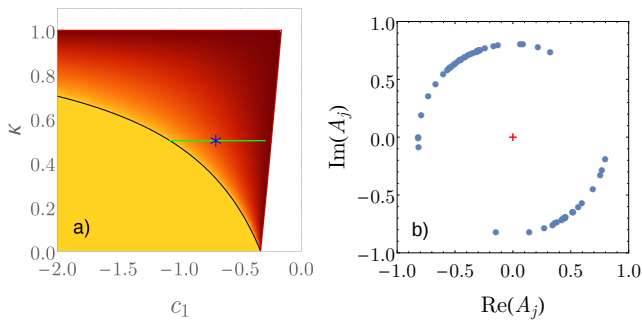}
	\caption{(a) Partial phase diagram of the MF-CGLE \eqref{MFCGLE} for $c_2=3$, showing the 
	domain of UIS and NUISs
	in the presence of an infinitesimal noise. In the yellow region UIS is stable, 
	while different NUISs are observed inside the other shaded region. 
	The color shading codes the 
	unevenness of the oscillator density
	through the value of $|Z_2|\equiv Q$.
	In the white region all incoherent states are unstable. The asterisk and the green line 
	indicate 
	the parameter values used in panel (b) and in Fig.~\ref{figZm}, respectively. (b) Snapshot of a random sample of $60$
	oscillators out of $N=300$, after a numerical simulation of $2\times10^6$ t.u. 
	where an independent white Gaussian noise 
	of intensity $D=10^{-6}$ along the real and imaginary parts of the $A_j$'s has 
	been added to remove the degeneracy among infinitely many neutrally stable NUISs.}
	\label{fignuis}
\end{figure}

In a broad region of parameter space the system \eqref{MFCGLE} settles into an incoherent state, i.e.~with zero mean field
$\overline{A}=0$. This does not specify the state of the system as it holds 
for a continuum of oscillator arrangements for $N>3$. 
The most prominent incoherent state is UIS, corresponding to oscillators located 
over a circle with uniformly distributed phases.
For the remaining incoherent states 
there is a lack of uniformity in the oscillator distribution and we use the acronym NUIS for them. 
Figure~\ref{fignuis}(a) shows a partial phase diagram for Eq.~\eqref{MFCGLE} for a specific value of $c_2=3$.
The UIS is observed in the light (yellow) shaded region at the left of the black solid line \cite{HR92,NK93}.
In the other shaded region a NUIS settles spontaneously (UIS is unstable). The asterisk in the
phase diagram indicates the parameter values for the snapshot of NUIS in Fig.~\ref{fignuis}(b). 
In this figure it is apparent that the oscillators are not evenly distributed, 
while $\overline A$, represented by a red cross, settles at the origin.

\section{($N+2$)-quasi phase reduction}
In this section we present our QPR method from $2N$ to $N+2$ degrees of freedom
for $N$ coupled $\lambda-\omega$ oscillators. The reduced system 
consists of $N$ phases plus 2 global degrees of freedom,
hence the name of QPR. 
The validity of the method requires a weak perturbation in the 
oscillators' motion, what holds in the neighborhood of the
incoherent states, irrespective of the coupling strength.

 \subsection{Coupling}
We will consider globally coupled identical oscillators:
	\begin{equation}\label{gentheeq}
	\dot{A_j}=f(A_j)+\kappa \, g(\mathcal{A}) ,
	\end{equation}
where $\kappa$ is a positive coupling constant, and $\mathcal{A}$ denotes one or more mean fields of the set $\mathcal{C}_A$, 
$\mathcal{A}\subseteq \mathcal{C}_A$. The set $\mathcal{C}_A$ of mean fields is:
	\begin{equation}
	\mathcal{C}_A=\left\{\overline{|A|^nA}\right\}_{n\in\mathbb{Z}}
	\cup \left\{\overline{|A|^nA^*}\right\}_{n\in\mathbb{Z}} .
	\end{equation}
Here $\overline{|A|^nA}=\frac{1}{N}\sum_{j=1}^N |A_j|^nA_j$ and  $^*$ stands 
for 
complex conjugation. Note that only the first harmonic in $\phi$ enters in the 
interaction. (The case with higher harmonics in the coupling
is discussed in Sec.~\ref{sechighharm}). Moreover, we demand the interaction function $g$
in Eq.~\eqref{gentheeq}
to vanish when the mean fields in the argument vanish, i.e.~$g({\cal A}=0)=0$.
Among the possible couplings the most preeminent one is diffusion, $g({\cal A})\propto \bar 
A$, as in the MF-CGLE \eqref{MFCGLE} introduced above \cite{HR92,NK93}. Other examples with nonlinear 
coupling 
as $g({\cal A})\propto  \bar A +b |\bar A|^2 \bar A$ 
and $g({\cal A})\propto \bar A + b \overline{| A|^2 A}$
have been considered in \cite{RP15} and \cite{Schmidt14}, respectively. It is also
important to notice that we do not exclude symmetry breaking terms in the coupling
such as $\mathrm{Re}(\bar A)$, similarly to \cite{break}.

\subsection{Preliminaries}

The first step of the analysis is to obtain
the evolution equation for the dynamics of the phases. 
Equation \eqref{gentheeq} in polar coordinates becomes
	\begin{subequations}
		\begin{eqnarray}\label{dotr}
		\dot{r}_j=\lambda(r_j)r_j 
+\kappa\operatorname{Re}\left[g(\mathcal{A})e^{-i\phi_j}\right] , \\
		\dot{\phi}_j=\omega(r_j) 
+\frac{\kappa}{r_j}\operatorname{Im}\left[g(\mathcal{A})e^{-i\phi_j}\right] .
		\end{eqnarray}
	\end{subequations}	
The phase dynamics is obtained through the change of variables in 
\eqref{iso}:
	\begin{equation}\label{pheqgen}
	\dot{\theta}_j=\Omega+\frac{\kappa}{r_j}	
\operatorname{Im}\left\{[1-ir_j\chi'(r_j)]g(\mathcal{A})e^{-i[\theta_j+\chi(r_j)
]}\right\}
	\end{equation}
here $\chi'(r_j)$ denotes the derivative of $\chi$ with respect to $r$ evaluated at
$r_j$. In order to reduce the dimensionality of the system we seek to 
remove the dependence on the radii $r_j$. 
Using Eq.~\eqref{dotr} we write the evolution equation for an infinitesimal
perturbation $\delta 
r_j$ off the limit cycle ($r_j=1+\delta r_j$):
	\begin{equation}\label{eqdotdr}
	\dot{\delta r_j}=\Lambda \delta 
r_j+\kappa\operatorname{Re}\left[g(\mathcal{A})e^{-i\theta_j}(1-i\chi_0\delta 
r_j)\right]+O(\delta r_j^2)
	\end{equation}
It is obvious that the oscillators will be in the proximity of the limit cycle whenever 
$\kappa | g(\mathcal{A})|\ll -\Lambda$. 
If this condition holds, 
we can set $r_j=1$ in \eqref{pheqgen}, obtaining thereby
the lowest order approximation:
	\begin{equation}\label{protophasered}
	\dot{\theta}_j=\Omega+\kappa
	\operatorname{Im}\left[(1-i\chi_0)g(\mathcal{A})e^{-i\theta_j}\right] .
	\end{equation}
This equation is not closed, as there are 
still dependences on the mean field(s) through $g(\mathcal{A})$.
	
\subsubsection{Small $\kappa$: Standard first-order phase reduction}

To put our work in context, and for later comparison, we note that traditional first order phase reduction 
assumes $\kappa\ll -\Lambda$, which automatically implies $\delta r_j\simeq 0$, as 
noted above. Therefore the mean fields in $g(\mathcal{A})$ can 
be approximated as
	\begin{equation}\label{AnAbarZ}
	\overline{|A|^nA}=\overline{r^{n+1}e^{i[\theta+\chi(r)]}}\simeq Z ,
	\end{equation}
where $Z\equiv\overline{e^{i\theta}}$ is the Kuramoto order parameter. Thus, 
at the lowest order, the coupling term will only depend on $Z$.
In this case we can make the replacement $g(\mathcal{A})\simeq\hat{\Gamma}(Z)$ 
in Eq.~\eqref{protophasered} obtaining
	\begin{equation}\label{tradpheq}
	\dot{\theta}_j=\Omega+\kappa
	\operatorname{Im}\left[(1-i\chi_0)\hat{\Gamma}(Z)e^{-i\theta_j}\right] .
	\end{equation}
This system of $N$ phase oscillators is the first-order phase reduction of
\eqref{gentheeq}. 
This reduction works poorly
if the coupling is not small; and even for asymptotically small coupling 
there are states of \eqref{gentheeq} not reproducible by Eq.~\eqref{tradpheq} such as NUIS 
(shown in Fig.~1(b)) or QPS.
Higher order terms proportional to $\kappa^2$, $\kappa^3$, etc.~can be 
incorporated 
into \eqref{tradpheq} removing degeneracies and 
extending the validity of the phase model with $N$ degrees of freedom 
\cite{leon19}. 
However, if the coupling is strong this procedure is either impractical 
(as the convergence rate of the series in
powers of $\kappa$ is not fast enough \cite{leon19}) 
or plain wrong (if the expansion in $\kappa$ is divergent). 

\subsection{Small $|g(\mathcal{A})|$: Quasi phase reduction of Eq.~\eqref{gentheeq}}

Incoherent states, the starting point of our analysis,
are configurations of the oscillators compatible with
${\cal A}=0$. Accordingly, in an incoherent state, 
each oscillator evolves as if it 
experienced no input from the rest of the population.
An ensemble of identical
oscillators may spontaneously settle into UIS or NUISs 
in wide regions of parameter space, see e.g.~Fig.~\ref{fignuis}.
Moreover, it is phenomenologically observed that there are also non-incoherent states in which 
strongly coupled  oscillators
behave as ``quasiphase oscillators'' \cite{CP19}, preserving their ordering on 
top
of a closed curve that evolves in time. This occurs, in particular, in globally 
coupled 
Stuart-Landau oscillators when UIS loses its stability
giving rise to a state called QPS which, after secondary
instabilities, yields pure collective chaos \cite{NK93,CP19}.

With the aim at describing the previous phenomena in a minimal way, we resort to Eq.~\eqref{protophasered}
since it already suggests that some kind of perturbative approach in small $g(\cal{A})$ is feasible 
in analogy to the small $\kappa$ approximation in standard phase reduction.
As Eq.~\eqref{protophasered} is not closed due to $g(\mathcal{A})$,
we are tempted to consider $g(\mathcal{A})$ as a new variable. This is not the 
best choice as the evolution equation cannot 
be generally closed in terms of $g(\mathcal{A})$. Instead, the complex variable 
$B=\overline{A}$ is the right choice, since, as shown below, any mean field 
$\overline{|A|^nA}$ can be approximately 
expressed in terms of $B$ and $Z$.
Assuming the proximity of the oscillators to their fiducial limit cycles, 
$r=1+\delta r$, we expand $\phi=\theta+\chi_0\delta r+O(\delta r ^2)$.
In this way the mean field $B$ is 
	\begin{equation}\label{eqBexpand}
	B=\overline{A}=\overline{r e^{i\phi}}\simeq\overline{(1+\delta r) 
e^{i(\theta+i\chi_0\delta r)}}\simeq Z+(1+i\chi_0)\overline{\delta r 
e^{i\theta}} .
	\end{equation}
Therefore, we can express the average $\overline{\delta r e^{i\theta}}$ in terms of $B$ and $Z$:
	\begin{equation}\label{eqdreth}
	\overline{\delta r e^{i\theta}}\simeq\frac{B-Z}{(1+i\chi_0)} ,
	\end{equation}
and apply this identity to all the other mean fields, obtaining a linear dependence
of $\overline{|A|^nA}$ on $B$ and $Z$:
	\begin{eqnarray}\label{AnAbar}
	\overline{|A|^nA}&=&\overline{r^{n+1}e^{i\phi}}\simeq 
Z+(n+1+i\chi_0)\overline{\delta r e^{i\theta}}\nonumber\\
	&\simeq& B+\frac{n}{1+i\chi_0}(B-Z) .
	\end{eqnarray}
With the previous equation any $g(\mathcal{A})$ can be approximated by a function of $Z$ and $B$:
	\begin{equation}\label{eqgBZ}
	g(\mathcal{A})\simeq\Gamma(Z,B)	.
	\end{equation}

Now the evolution of $B$ is obtained averaging 
\eqref{gentheeq}
over the whole population, Namely,
	\begin{equation}\label{Bdyn}
	\dot{B}=\frac{1}{N}\sum_{k=1}^{N}\dot{A_k}=\overline{f(A)}+\kappa ~ 
g(\mathcal{A}).
	\end{equation}
The term $\overline{f(A)}$ is calculated using
Eqs.~\eqref{fexpan}, \eqref{nfn}, and \eqref{AnAbar}:
	\begin{equation}\label{eqfBZ}
	\overline{f(A)}\simeq i\Omega B+\Lambda\left(B-Z\right)
	\end{equation}

Finally, replacing Eqs.~\eqref{eqgBZ} and \eqref{eqfBZ} into  
Eqs.~\eqref{protophasered} and \eqref{Bdyn} we obtain the $(N+2)$-QPR
of the globally coupled oscillator system defined by Eq.~\eqref{gentheeq}:
\begin{subequations}\label{quasiredtheo}
	\begin{eqnarray}
	\dot{\theta}_j&=& 
\Omega+\kappa\operatorname{Im}\left[(1-i\chi_0)\Gamma(Z,B)e^{-i\theta_j}\right] 
\label{phaquasiredtheo}\\
	\dot{B}&=&i \, \Omega \, B+\Lambda\left(B-Z\right)+\kappa \,\Gamma(Z,B)  
\label{bquasiredtheo}
	\end{eqnarray}
\end{subequations}
These equations are the main result of this paper. Some important remarks follow.

\subsubsection{Remarks on the $(N+2)$-QPR, Eq.~\eqref{quasiredtheo}}

The QPR that transforms \eqref{gentheeq} into \eqref{quasiredtheo} 
entails a
drastic decrease in the number of
degrees of freedom from $2N$ to $N+2$: $N$ phases plus a complex 
 collective variable $B$. 
In contrast to standard phase reduction, there is an extra complex variable $B$.
This is the key ingredient to make the strong coupling amenable to analysis, while preserving
the population of phase oscillators.
The theory is consistent since QPR \eqref{quasiredtheo} boils down to the 
standard phase reduction \eqref{tradpheq}  in the $\kappa\rightarrow 0$ limit. 
To see this, set $\Omega=0$ in \eqref{quasiredtheo} by going to a rotating frame 
$(\theta_j',B')=(\theta_j-\Omega t,B e^{-i\Omega t})$ if necessary,
and note that $B(t)\to Z(t)$ as $\kappa\rightarrow 0$ in 
Eq.~\eqref{bquasiredtheo}.
In this way,  Eq.~\eqref{phaquasiredtheo} reduces to \eqref{tradpheq} since 
$\Gamma(Z,Z)=\hat{\Gamma}(Z)$, cf.~Eqs.~\eqref{AnAbarZ} and \eqref{AnAbar}.

Equation \eqref{quasiredtheo} can be regarded as a population of phase 
oscillators coupled through a sort of external medium $B$. 
Indeed, a similar model is obtained applying ordinary phase reduction (assuming weak coupling) 
to a model of `dynamical quorum sensing' in which oscillators are coupled through a medium with 
intrinsic dynamics \cite{schwab12}. 
Here, in sharp contrast, there is no `medium' in the 
original system \eqref{gentheeq}, instead
QPR endows the mean field with
a virtual dynamical equation.

An important feature of Eq.~\eqref{quasiredtheo} (as a consequence of the approximations \eqref{eqBexpand} and \eqref{eqdreth}) is that it is
a quasi-integrable model that can be analyzed within the framework of the 
Watanabe-Strogatz theory \cite{WS94,PR11}.
Given a particular initial condition there are $N-3$ constants of motion 
determining the fate of the system. This  
degeneracy of the model is not present in \eqref{gentheeq}. Still, the system in 
Eq.~\eqref{quasiredtheo} 
is useful at least because of two reasons: (i) we can use it to determine the stability 
(boundary)
of incoherent states analytically, see next sections; and (ii) it is the 
starting point for higher-order QPR, see Sec.~\ref{sec_next}.

\subsubsection{Stability of incoherent states}
\label{sec:theory}

Equation~\eqref{quasiredtheo} is the QPR of model \eqref{gentheeq}, irrespective of the number $N$ of oscillators.
In this section we take the thermodynamic limit ($N\to\infty$) 
and analyze the stability boundary of the incoherent states. 
The analysis requires defining a density $\rho$ such that 
$\rho(\theta,t)d\theta$ is the fraction of oscillators 
with phases between $\theta$ and $\theta+d\theta$ at time $t$. Additionally, we impose the 
normalization condition $\int_{0}^{2\pi}\rho(\theta,t)d\theta=1$.
The Kuramoto order parameter is now $Z=\int_{0}^{2\pi}\rho(\theta,t) e^{i\theta} d\theta$.
The oscillator density $\rho$ obeys the continuity equation because of the 
conservation of the number of oscillators:
\begin{equation}\label{conteq}
\partial_t  \rho(\theta,t) + \partial_\theta[ v(\theta) \rho(\theta,t)] =0 .
\end{equation}
This is a nonlinear equation since $v=\dot{\theta}$ depends on $\rho$. 

According to Eq.~\eqref{AnAbar}, all $\overline{|A|^nA}$ are linear combinations of $B$ and $Z$. 
Therefore, all states with $B=Z=0$ are incoherent states since $\Gamma(0,0)=0$.
Obviously, there are infinitely many phase densities compatible with
$Z=0$, which rotate uniformly: $\rho_{incoh}(\theta,t)=\rho_{incoh}(\theta-\Omega t)$.
Notably, it will be shown below that not all incoherent states become unstable simultaneously.

The analysis proceeds introducing the Fourier expansion of $\rho$:
\begin{equation} \label{fourier}
\rho(\theta,t)=\frac{1}{2\pi}\sum_{m=-\infty}^\infty\rho_m(t)e^{-im\theta}
\end{equation}
with coefficients $\rho_0=1$ and $\rho_{-m}=\rho_m^*$. Inserting \eqref{fourier} into \eqref{conteq},
and noting that $Z=\rho_1$, we may rewrite our model \eqref{quasiredtheo} in Fourier space:
	\begin{subequations}\label{evol_modes}
\begin{eqnarray}
\dot{\rho}_m=i m\Omega \rho_m+ \frac{m 
\kappa}{2}\bigg[(1-i\chi_0)\Gamma(\rho_1,B)\rho_{m-1} \nonumber\\
-(1+i\chi_0)\Gamma^*(\rho_1,B)\rho_{m+1}\bigg] \\
	\dot{B}=i \Omega B+\Lambda\left(B-\rho_1\right)+\kappa \Gamma(\rho_1,B) .
	\end{eqnarray}
	\end{subequations}
In the light of these equations 
it becomes apparent the existence of an infinite set of incoherent solutions 
characterized by $\rho_1=B=0$, and $\rho_{m\ge2}=\hat{\rho}_m e^{im\Omega t}$ with arbitrary $\hat{\rho}_{m\ge2}$. 
We distinguish between UIS, corresponding to $\hat{\rho}_{m\ne0}=0$, 
and the remaining set of NUISs. 

The linear stability of (N)UIS is determined 
considering the evolution of infinitesimal perturbations of the form $\rho_m=(\hat{\rho}_m+\delta 
\rho_m)e^{im\Omega t}$ and $B=\delta Be^{i\Omega t}$. The linearization of Eq.~\eqref{evol_modes} turns out to be:
	\begin{subequations}\label{genthemodes}
		\begin{eqnarray}\label{genthemodesrho}
		\dot{\delta\rho_m}=\frac{m 
\kappa}{2} \left[
(1-i\chi_0)e^{-i\Omega}\hat{\rho}_{m-1}\vec{\nabla}
\Gamma\cdot \vec{\delta} \right.\nonumber\\		
\left. -(1+i\chi_0)e^{i\Omega}\hat{\rho}_{m+1}\vec{\nabla}\Gamma^*\cdot 
\vec{\delta} \right] ,\\
		\dot{\delta B}=i\Omega\delta B+\Lambda(\delta 
B-\delta\rho_1)+\kappa\vec{\nabla}\Gamma\cdot \vec{\delta} .
		\end{eqnarray}
	\end{subequations}
The right-hand sides of these equations only include perturbations in the subspace spanned by $\rho_1$ and $B$;
note the shorthand notation $\vec{\delta}=(\delta \rho_1,\delta \rho_1^*,\delta B,\delta B^*)^T$, and
the gradients $\vec\nabla\Gamma$ defined in this subspace and evaluated at $\rho_1=B=0$.
We then have an infinite set of vanishing eigenvalues corresponding to
eigenvectors with $\delta B=\delta \rho_{1}=0$ 
\footnote{The naive expectation is that these neutral modes should decay to zero under arbitrarily weak
noise, as observed in the UIS of the MF-CGLE \cite{NK93}. Actually, this is not necessarily the case,
 as a specific example in the next section shows.}

Hence, according to Eq.~\eqref{genthemodes}, the relevant infinitesimal 
instabilities develop in the subspace 
spanned by $\rho_1$ and $B$. We are led to analyze the $4\times4$ 
Jacobian matrix ruling the dynamics of ${\delta B}$ and ${\delta\rho_1}$. 
In this Jacobian only the second
mode $\hat{\rho}_2$ (and $\hat{\rho}_2^*$) 
is present.  Moreover, it can be shown that 
the stability of all incoherent states can 
be classified by the value of the amplitude $|\hat{\rho}_2|=Q$.
This result was already proved in a particular case \cite{HR92,chabanol97},
 but QPR shows that it is 
a general property of the coupling via the mean fields in $\mathcal{C}_A$. 

Finally, we want to stress that the stability boundaries of (N)UIS
obtained from 
\eqref{genthemodes} exactly match those of the original system \eqref{gentheeq}. 
The reason is that QPR is asymptotically valid
in the limit $g(\mathcal{A})\to0$, i.e.~where the instabilities take place.


\section{Quasi phase reduction for Stuart-Landau oscillators}

In this work we address populations of Stuart-Landau oscillators in detail.
Reduction via QPR for other $\lambda-\omega$ oscillators is worked out likewise. 

\subsection{Linear coupling: mean-field complex Ginzburg-Landau equation}
A simple system to illustrate and test our previous findings is the MF-CGLE presented in Sec.~\ref{secCGLE}. 
Written as in \eqref{MFCGLE} the values of $\Lambda=-2$ and $\chi_0=c_2$ remain those indicated in Sec.~\ref{SL}, and
given  that $g(\mathcal{A})=(1+ic_1)\overline{A}$, it is 
straightforward to obtain
$\Gamma(Z,B)=(1+ic_1)B$. Hence, the quasi 
phase reduced model \eqref{quasiredtheo} becomes:
	\begin{subequations}\label{MFCGLEmod}
		\begin{eqnarray}
		\dot{\theta_j}&=&\kappa\, \eta \, |B| \sin(\Upsilon-\theta_j+\alpha)\label{MFCGLEph} \\
		\dot{B}&=&-2\left(B-Z\right)+\kappa\, (1+ic_1) B \label{MFCGLEmf} 
		\end{eqnarray}
	\end{subequations}
where $B=|B|e^{i\Upsilon}$, $\eta\equiv\sqrt{(1+c_2^2)(1+c_1^2)}$, 
$\alpha\equiv\arg[1+c_1 c_2+(c_1-c_2)i]$. Equation \eqref{MFCGLEmod}
is similar to the Kuramoto-Sakaguchi model \cite{Kur84}, but with the phase oscillators
coupled through $B$ instead of $Z$. Only in the limit $\kappa\to0$, $B$ approaches
$Z$ and the standard first-order phase reduction is recovered \cite{NK93}.

\subsubsection{Numerical results: Transient dynamics}

\begin{figure}
 	\includegraphics[width=\linewidth]{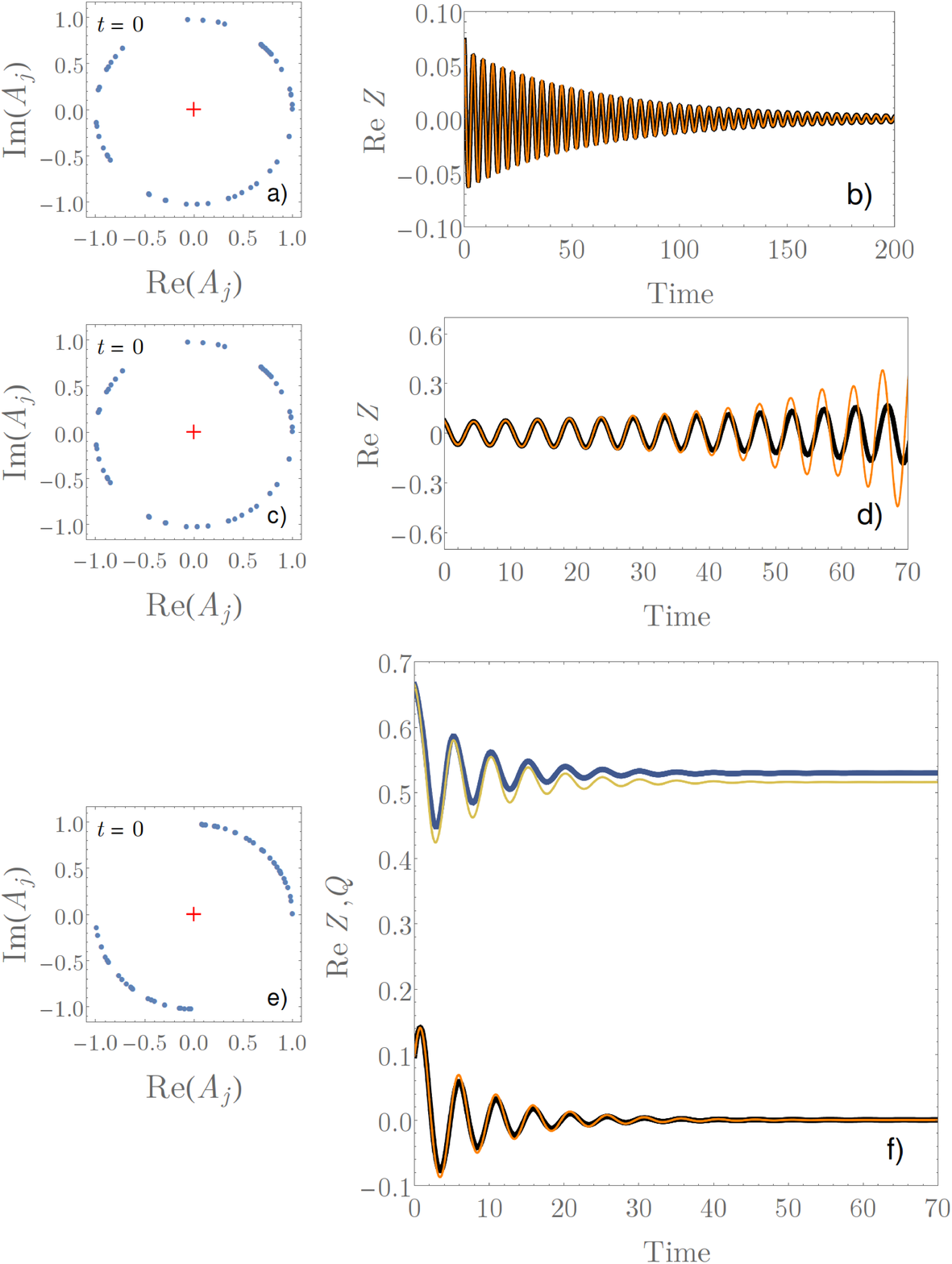}
  	\caption{Time series of the real part of Kuramoto order parameter $Z$ 
(b,d,f) for the MF-CGLE \eqref{MFCGLE} and its QPR  \eqref{MFCGLEmod} with $N=50$ 
depicted by black and orange lines, respectively. In (f) the modulus 
of second Kuramoto-Daido order parameter $Q$ is also depicted, 
with dark blue and yellow colors for the MF-CGLE and its QPR, respectively. 
In panels (b) and (d) the 
oscillators are initially distributed randomly over the unit circle, as shown in 
panels (a) and (c). In (f) the oscillators' phases $\theta_j$ are 
randomly initialized over the interval $[0,\frac{\pi}{2}]\cup[\pi,\frac{3\pi}{2}]$, 
as shown in panel (e). Accordingly, the system is near the a NUIS  with $B=Z\simeq0$ and $Q\simeq 
0.66$. The parameters chosen are $c_2=3$, $\kappa=0.5$ and $c_1=-1.1$ in (b) 
where UIS is stable and $c_1=-1$ in (d) and (f) where UIS is unstable but 
NUIS with $Q>\frac{1}{3\sqrt{2}}\simeq0.24$ are not.}s
 	\label{transient}
 \end{figure}

To confirm the correctness of our approach we compare the 
transient behavior of the MF-CGLE \eqref{MFCGLE}
with its QPR \eqref{MFCGLEmod}.
We track the evolution of the mean field $Z=\overline{e^{i\theta}}$ for both systems
near incoherent states, noting that for the
MF-CGLE $Z$ is $\overline{e^{i(\phi-c_2\ln r)}}$.
In Fig.~\ref{transient}(a-d) we initialized $N=50$ oscillators randomly on 
the unit circle, i.e.~near the UIS. The 
parameters used in Figs.~\ref{transient}(a,b) and \ref{transient}(c,d) 
correspond to stable and unstable UIS, respectively. 
The stability properties of UIS, decay/growth rate and oscillation frequency,
are perfectly captured by the QPR equations \eqref{MFCGLEmod}. 
As expected, in Fig.~\ref{transient}(d) after a certain time interval, 
the mean field $|Z|$ 
grows too large and the QPR equations become inaccurate (the MF-CGLE 
approaches a saddle quasiperiodic partial synchrony and eventually decays to a NUIS). 
In Figs.~\ref{transient}(e) and \ref{transient}(f) we show that 
QPR also gives a good 
description of NUISs. With the same parameters that in Fig.~\ref{transient}(d),
the oscillators were randomly set in the phase interval 
$[0,\frac{\pi}{2}]\cup[\pi,3\frac{\pi}{2}]$ of the unit circle. In this way $B=Z\simeq0$ but 
$Q=\left|\overline{e^{i2\theta}}\right|\simeq 
2/\pi$. We can see in Fig.~\ref{transient}(f) 
that, as time evolves, $Z$ decays to zero but $Q$ converges to a nonzero constant value 
because UIS is unstable, but NUISs with large enough $Q$ values are not.

\subsubsection{Uniform incoherent state}
A closed formula for the stability boundary of the UIS was already found in \cite{HR92,NK93}, so here 
we just wish to evidence how QPR permits to obtain it in a simple way.
As mentioned above only the evolution of $\delta\rho_1$ and $\delta B$
must be taken into account in Eq.~\eqref{genthemodes}. As
${\rho}_{2}=0$ in the UIS, we get:
	\begin{equation}
	\frac{d}{dt} \begin{pmatrix}\delta \rho_1 \\ \delta B \end{pmatrix} =
	\left(
	\begin{array}{cc}
0&	\frac{\kappa\eta}{2}e^{i\alpha} \\  
-2  &	  -2+\kappa(1+ic_1) 
	\end{array}
	\right) 
	\begin{pmatrix} \delta\rho_1\\ \delta B \end{pmatrix}  , \nonumber
	\end{equation}
The characteristic equation is: 
	\begin{equation}
	P_2(\lambda)=\lambda^2+(2-\kappa-i\kappa c_1)\lambda-\kappa \eta 
e^{i\alpha}=0 . \nonumber
	\end{equation}
The locus of the (oscillatory) instability is determined imposing $\lambda=i\Omega_c$. The 
critical coupling $\kappa_0$ satisfies:
	\begin{equation}\label{UIS}
	\kappa_0(\kappa_0-1)c_1^2-4(\kappa_0-1) c_1 
c_2+\kappa_0c_2^2+(\kappa_0-2)^2=0 ,
	\end{equation}
in agreement with \cite{HR92,NK93} (be aware of the different parametrizations in each work).

\subsubsection{Non-uniform incoherent state}
The stability boundary of each NUIS is determined 
through the evolution of $\delta \rho_1$ and $\delta B$. Inserting the specific value of $\hat\rho_2=Q\le1$ 
into Eq.~\eqref{genthemodes} we get:
	\begin{subequations}
		\begin{eqnarray}
		\delta\dot{\rho_1}&=\frac{\kappa\eta}{2}\left(e^{i\alpha}\delta 
B-Qe^{-i\alpha}\delta B^*\right) \\
		\delta\dot{B}&=-2\left(\delta 
B-\delta\rho_1\right)+\kappa(1+ic_1) \delta B 
		\end{eqnarray}
	\end{subequations}
The associated characteristic polynomial of fourth degree is:
	\begin{equation}
	P_4(\lambda)=\lambda^4+a_1\lambda^3+a_2\lambda^2+a_3\lambda+a_4 . \nonumber
	\end{equation}
Although the zeros cannot be computed, the Routh-Hurwitz criterion 
\cite{Gantmacher89} allows to know if there is at least one root with 
nonnegative real part. For the fourth order polynomial $P_4(\lambda)$
all roots have negative real parts if and only if $a_i>0$ and 
$a_1a_2a_3-a_1^2a_4-a_3^2>0$. This criterion gives five conditions for 
the stability of a particular ``$Q$-NUIS'':
	\begin{subequations}\label{NUISboundary}
	\begin{multline}\label{NUISboundarya}
	\kappa_Q(\kappa_Q-1)c_1^2-4(\kappa_Q-1)c_1c_2
	+\kappa_Q c_2^2+(\kappa_Q-2)^2 ,\\
	+\frac{\kappa_Q^2(1+c_1^2)(1+c_2^2)}{[(2-\kappa_Q)^2+c_1^2]}Q^2>0
	\end{multline}
	\begin{equation}\label{NUISboundaryb}
	4-2\kappa_Q(3 c_1c_2)+(1+c_1^2)\kappa_Q^2>0 ,
	\end{equation}
	\begin{equation}
	\kappa_Q<2 ,
	\end{equation}
	\end{subequations}
plus two other inequalities that are always fulfilled. Equations 
\eqref{NUISboundary} are precisely the exact $Q$-dependent NUIS stability 
boundaries of \eqref{MFCGLE} \cite{chabanol97}.

\subsubsection{The effect of arbitrarily weak noise}

\begin{figure}
	\includegraphics[width=\linewidth]{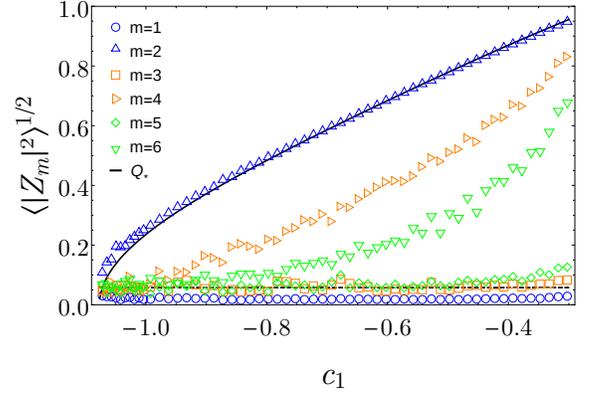}
	\caption{ 
	Root mean square $\langle|Z_m|^2\rangle^{1/2}$ of the Kuramoto-Daido order 
	parameters $Z_m=\frac{1}{N}\sum_{k=1}^{N}e^{im\theta_k}$ ($m=1,\ldots,6$)
	along the green line in Fig.~\ref{fignuis}(a) ($c_2=3,\kappa=0.5$). 
	The black line is $Q_*$, the theoretically predicted value of $|Z_2|$, while the horizontal
	dashed line at $1/\sqrt{N}$ (roughly) indicates the upper expected value of the statistical fluctuations for
	a vanishing $Z_m$ in the thermodynamic limit.
	The simulations were carried out with $N=300$ oscillators, under 
	independent white Gaussian noises of intensity $D=10^{-6}$ along the real and imaginary parts of the $A_j$'s.}
	\label{figZm}
\end{figure}

A particular (N)UIS may be either unstable or neutrally stable, but not asymptotically stable.
Thus, in the MF-CGLE a continuum of neutrally stable incoherent states coexist in regions of parameter space. 
 Hence, the question is the selective effect of arbitrarily weak noise.
The color shading in the phase diagram of Fig.~\ref{fignuis}(a) has been made from Eqs.~\eqref{UIS} and \eqref{NUISboundary}
under the assumption that
the system adopts a phase density with $Q=Q_*$, where $Q_*$ is the smallest $Q$ value
among all neutrally stable NUISs. 
Indeed, a neutral UIS is attracting in the presence of weak noise, see \cite{NK93}.
As may be seen in Fig.~\ref{figZm}, in the region where UIS is unstable the values of 
$Q$ observed match almost perfectly with $Q_*$, depicted by a black solid line.

Less intuitive is the behavior of the remaining modes, $Z_m$ ($m>2$), that are irrelevant in the
stability analysis. According to our Fig.~\ref{figZm}, in the UIS region all $Z_m$ go to zero,
while in the NUIS region this is only the case for odd $m$ index. The even modes
grow as $Q$ increases. The last NUIS to destabilize is $Q=1$
and corresponds to two equally populated point clusters in antiphase,
i.e.~a bi-delta phase density ($Z_2=Z_4=Z_6=\cdots=e^{i\xi}$).

\begin{figure*}
	\begin{minipage}[t]{0.98\linewidth}
		\includegraphics[width=\textwidth]{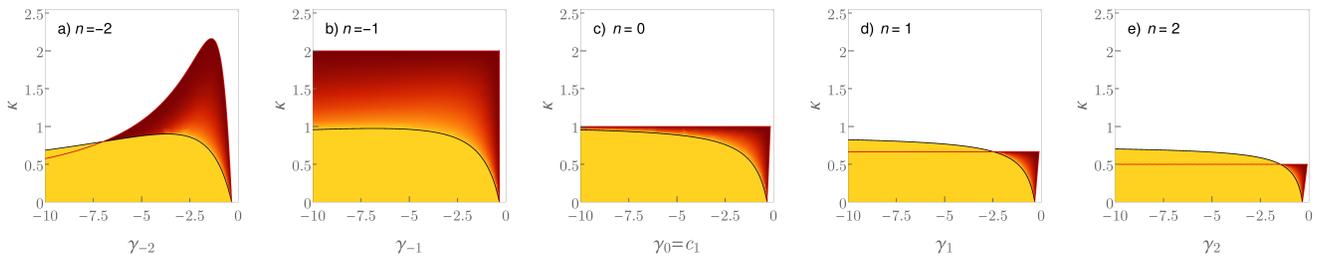}
	\end{minipage}
	\caption{Partial phase diagrams for populations of Stuart-Landau oscillators 
with nonisochronicity parameter $c_2=3$, coupled through $g({\cal A})=(1+i\gamma_n)\overline{|A|^nA}$
with $n=-2,1,0,1,2$, in panels (a)-(e). The stability boundaries of UIS and the NUIS with $Q=1$
are depicted by black and red lines, respectively. The UIS is stable in the yellow region.
In the shaded region the intensity of the red color indicates $Q_*$ (the smallest
$Q$ value among the non-unstable NUISs).}
	\label{Fignlphasediagram}
\end{figure*}


\subsection{Nonlinear coupling, $g({\cal A})\propto\overline{|A|^nA}$}

Recent papers by Schmidt, Krischer and coworkers \cite{Schmidt14,schmidt15} have studied 
chimera states in the MF-CGLE with
an extra coupling term proportional to $\overline{|A|^2A}$. Here, instead of 
embarking on the exploration of the high-dimensional parameter space
{of that system,
each coupling of the form $\overline{|A|^nA}$
($n\in\mathbb Z$) 
is analyzed separately.}
Thus, the systems under consideration are: 
	\begin{equation}\label{NLSL}
	\dot{A_j}=(1+i c_2) (1-|A_j|^2)A_j+\kappa (1+i \gamma_n) 
\overline{|A|^nA} ,
	\end{equation}
where the parameter $\gamma_n$ is a real constant. In the particular case $n=0$, Eq.~\eqref{NLSL}
becomes the MF-CGLE \eqref{MFCGLE} and $\gamma_0=c_1$, accordingly.

Deriving the QPR of \eqref{NLSL} requires calculating the function $\Gamma(Z,B)$.
Using \eqref{AnAbar} with $\chi_0=c_2$ the result is straightforward: 
	\begin{equation}\label{Gamman}
\Gamma_n(Z,B)=(1+i\gamma_n)\left[B+\frac{n}{1+i c_2} 
(B-Z)\right] ,
	\end{equation}
where the subscript $n$ is used to indicate the dependence on the
specific coupling considered. 
Finally, inserting $\Gamma_n$ into \eqref{quasiredtheo} 
we obtain the QPR of \eqref{NLSL}:
	\begin{subequations}\label{NLSLred}
		\begin{eqnarray}
		\dot{\theta}_j&=&\kappa\eta_{B} |B|\sin 
(\Upsilon-\theta_j+\alpha_{B}) -\kappa \eta_{R} R\sin(\Psi-\theta_j+\alpha_{R}) \nonumber\\& \label{casiKS}\\
\dot{B}&=&\Lambda\left(B-Z\right)+\kappa \, \Gamma_n(Z,B)
		\end{eqnarray}
	\end{subequations}
where we have defined 
$\eta_{B}e^{i\alpha_{B}}=\frac{(1-i c_2)(1+i\gamma_n)(1+n+i c_2)}{1+ic_2}
$, $\eta_{R} e^{i\alpha_{R}}=\frac{n(1-i c_2)(1+i\gamma_n)}{1+i c_2}$ and 
$Z=R e^{i\Psi}$.

Prior to determining the exact stability boundaries of (N)UIS from Eq.~\eqref{NLSLred}, 
let us see what the standard first-order phase reduction predicts.
For this, we take the limit $\kappa\to0^+$,
observing that $B$ collapses into $Z$, and Eq.~\eqref{casiKS} 
becomes the Kuramoto-Sakaguchi model in Eq.~\eqref{tradpheq}. The crossover
from perfect synchrony to incoherence is given by the Benjamin-Feir-Newell criterion ($1+c_1c_2=0$)
in the diffusive case, which now generalizes to:
	\begin{equation}\label{genrealizedbencrit}
	1+\gamma_nc_2=0 ,
	\end{equation}
by virtue of Eq.~\eqref{AnAbarZ}.		
Therefore, in a phase diagram of system \eqref{NLSL} including 
the $\kappa$ axis, 
the stability boundaries of (N)UIS are expected
to emanate from $\gamma_n=-c_2^{-1}$
at $\kappa=0$.
	
The exact stability boundaries of UIS 
and NUISs in the thermodynamic limit are obtained from \eqref{NLSLred} as explained in previous sections. The 
stability boundary of UIS is:
	\begin{multline}
	\kappa_0[4 (c_2-\gamma_n )^2+\gamma_n  (n+2) (\gamma_n -c_2) [\kappa_0  
(n+2)-4)]\\
	+(\gamma_n  c_2+1) (\kappa_0  (n+2)-4)^2=0
	\end{multline}
We can see in the limit $\kappa\to0$ we recover \eqref{genrealizedbencrit}. The 
stability boundary of a $Q$-dependent NUIS can be computed as was done in the linear coupling 
case; the interested reader can find its expression in the Appendix.

In Fig.~\ref{Fignlphasediagram} the stability boundaries lines of UIS and NUIS 
are depicted for five different values of $n=-2,-1,\ldots,2$. 
Taking $n=0$ in panel (c)  as the reference case, 
we see that augmenting $n$ shrinks the region of incoherence.
On the contrary, for $n=-1$ stable NUISs reach larger $\kappa$ values, 
while the UIS region remains mostly unchanged. The boundaries for 
other negative $n$ values are similar to those for $n=-2$ in Fig.~\ref{Fignlphasediagram}(a). 
It is interesting
that for all $n$ 
values there are regions in parameter space where UIS is unstable,
but certain NUISs are not. This means that, at least for certain initial conditions,
the system may spontaneously converge to a NUIS. According to our numerical simulations,
and as reasoned above, under weak noise the non-unstable 
NUIS with the smallest $Q$ value is observed.
In addition, save for $n=-1,0$, there are also regions 
for small enough $\gamma_n$  
where UIS is the last incoherent state to become unstable.

We want to remark that all the stability boundaries calculated are the exact 
results for \eqref{NLSL} and their correctness has been 
numerically checked using an ensemble of $N=100$ oscillators (not shown). To our knowledge, only the case of the MF-CGLE
had been solved so far \cite{HR92,chabanol97}. We believe using QPR \eqref{NLSLred} is the
most effective method for computing these boundaries.


\subsection{Other couplings}

In this subsection we want to make some comments on the couplings where
the $(N+2)$-QPR scheme presented so far can be applied.

It is straightforward to consider a combination of nonlinear couplings such as:
	\begin{equation}
g(\mathcal{A})=\sum_{n=-\infty}^{\infty}\sigma_n\overline{|A|^nA} 
+\mu_n\overline{|A|^nA^*} \nonumber
	\end{equation}
	with complex $\sigma_n$ and $\mu_n$.
In this case, $\Gamma(Z,B)$ is simply a sum over terms like the bracketed part in the right-hand side
of Eq.~\eqref{Gamman} and their complex conjugates.

Other nonlinear coupling considered in \cite{PR09,RP15}:
	\begin{equation}
	   g(\mathcal{A})=(\epsilon_1+i \epsilon_2) \overline{A}-\sigma 
(\eta_1+i \eta_2)\left|\overline{A}\right|^2\overline{A} , \nonumber
	\end{equation}
can be treated analogously to other couplings. Nonetheless, the stability boundaries of 
UIS and NUISs in this case are the same that those for Eq.~\eqref{MFCGLE} because the 
nonlinear term is negligible if $|B|=|\overline{A}|\ll1$.

Our QPR approach does not exclude systems with 
nonlinear delayed feedback and/or 
couplings such as  $h(A_j)g(\mathcal{A})$ (provided $h$ has polar symmetry 
$h(A_je^{i\phi})=e^{i\phi}h(A_j)$ \footnote{This condition
permits to calculate the equation for $\dot B$ using Eq.~\eqref{AnAbar}.}), similar to those studied in \cite{PHT05}.

Finally, the case of a scalar coupling like $g({\cal A})\propto\operatorname{Re}(\overline{A}) 
\propto\bar A +\bar A^*$, is particularly simple, as QPR may be further reduced to only one
real-valued global variable, i.e.~$N+1$ degrees of freedom in total.

\section{Exploring the next order of QPR}
\label{sec_next}

As it occurs with standard first-order phase reduction, extending the theory to the next order
in the QPR scheme is not a trivial task. For QPR, a systematic expansion is even more troublesome
as there is not a small coupling parameter, but a small field $g({\cal A})$. 
This should be the goal of future work, but we think it may be instructive to  
pinpoint the difficulties, as well as to
examine the workable limit of small coupling.  

The first step is to expand Eq.~\eqref{pheqgen} to order $\delta r_j$. We obtain in this way
an augmented version of Eq.~\eqref{protophasered}:
	\begin{eqnarray}\label{eqthexpand2}
	\dot{\theta}_j&=&\Omega+\kappa
\operatorname{Im}\left[(1-i\chi_0)g(\mathcal{A})e^{-i\theta_j}\right] \nonumber\\
&-&\kappa \operatorname{Im}\left\{\left[1+\chi_0^2+i(\chi_0+\chi_1)\right]
g(\mathcal{A})e^{-i\theta_j}\right\} \delta r_j ,
	\end{eqnarray}
where $\chi_1=\frac{d^2\chi(r)}{dr^2}|_{r=1}$. 
The deviation from the reference radius
$\delta r_j$ evolves in time as dictated by
Eq.~\eqref{eqdotdr}, which is coupled to $\theta_j$ and to the mean field $\cal A$.
It is not obvious how to proceed next since $\cal A$ is not static.

\subsection{Small $\kappa$}
Inspecting Eq.~\eqref{eqdotdr} we realize that
if $\kappa$ is small then $\delta r_j(t)$ adjusts quickly to the current mean field:
	\begin{equation}\label{eqdeltar2}
	\delta 
r_j=-\frac{\kappa}{\Lambda}\operatorname{Re}\left[g({\cal A})e^{-i\theta_j} 
\right] + O(\kappa^2)
	\end{equation}
We can insert \eqref{eqdeltar2} into \eqref{eqthexpand2} to obtain the phase 
equation up to $O(\kappa^2)$ 
	\begin{eqnarray}\label{dthetak2}
	&&\dot{\theta}_j=\Omega+\kappa \operatorname{Im}\left[(1-i\chi_0)g 
e^{-i\theta_j}\right] \\
&+&\frac{\kappa^2}{\Lambda}  \left\{ \frac{1+\chi_0^2}{2} \operatorname{Im} 
\left(g^2e^{-i2\theta_j}\right)
	+(\chi_0+\chi_1)
\left[\operatorname{Re}\left(g e^{-i\theta_j}\right)\right]^2 \right\} . \nonumber
\end{eqnarray}
With the new term, proportional to $\kappa^2$, the Watanabe-Strogatz theory \cite{WS94}
cannot be applied. This is not a surprise, since 
the original model is not quasi-integrable.
To proceed with the analysis, function $g$ has to be written in terms of the 
the mean fields $Z$, $B$, and maybe others. To the lowest order we 
simply { adopt} 
the function $\Gamma(Z,B)$ obtained above.

\subsubsection{Mean-field complex Ginzburg-Landau equation}
Let us see how Eq.~\eqref{dthetak2}
applies to the particular case
of the MF-CGLE, Eq.~\eqref{MFCGLE}.
As the unit oscillator is the Stuart-Landau oscillator,
we insert $\chi_0=-\chi_1=c_2$ into Eq.~\eqref{dthetak2}.
Moreover, we keep the evolution for $B$ as before.
This results in an extended QPR model:
	\begin{subequations}\label{MFCGLEmod2nd}
		\begin{eqnarray}
		\dot{\theta_j}&=&\kappa\eta |B| 
\sin(\Upsilon-\theta_j+\alpha)-
 \frac{\kappa^2\eta^2|B|^2}{4}
\sin[2(\Upsilon-\theta_j)+\beta] , \nonumber \\ \label{MFCGLEph2nd} \\
		\dot{B}&=&-2\left(B-Z\right)+\kappa(1+ic_1) B ,
		\end{eqnarray}
	\end{subequations}
here $\beta=\arg(1-c_1^2+2i c_1)$.

\begin{figure}
 \includegraphics[width=\linewidth]{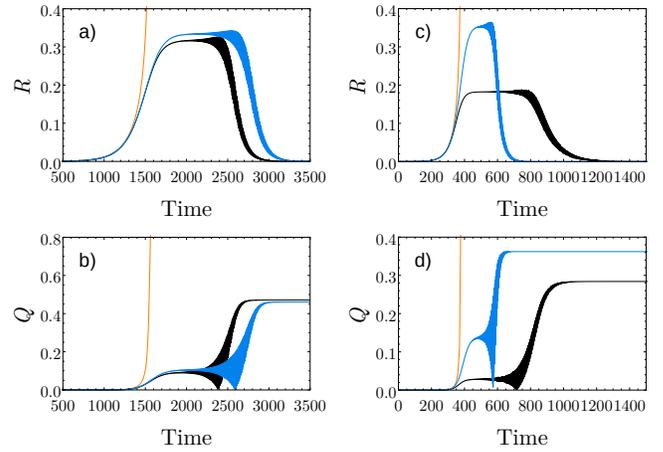}
 \caption{Time evolution of the MF-CGLE \eqref{MFCGLE} in black, its ($N+2$)-QPR \eqref{MFCGLEmod} 
 in orange, and the extended ($N+2$)-QPR \eqref{MFCGLEmod2nd} in blue.
 Panels (a,c) show the magnitude of the Kuramoto order parameter $R(t)=|Z(t)|$ and  
 (b,d) the magnitude of the second Kuramoto-Daido order parameter $Q(t)=|Z_2(t)|$.
 Two sets of parameters are used:
(a,b) $c_1=-0.5$, $\kappa=0.2$; (c,d) $c_1=-1$, $\kappa=0.5$.
Parameter $c_2=3$ in all panels.}
 	\label{QPStransient}
 \end{figure}

Next we test \eqref{MFCGLEmod2nd} by comparing with numerical simulations.
We select constants $c_1$, $c_2$, and $\kappa$ such that UIS and full synchrony
are both unstable, but NUISs with $Q$ above a certain value have not
destabilized. As observed in Ref,~\cite{leon19}, 
for small and moderate $\kappa$ values there is a heteroclinic connection between UIS and a saddle QPS. 
Recall that, in a QPS state
the oscillator density rotates uniformly 
(as $Z$, $Z_2$, etc, accordingly), but each individual oscillator exhibits quasiperiodic motion.
For the numerical test in Fig.~\ref{QPStransient} we initialize $N=100$  Stuart-Landau oscillators 
randomly in the unit circle for the full model \eqref{MFCGLE}, as well as
the ($N+2$)-QPR \eqref{MFCGLEmod},
and the extended ($N+2$)-QPR \eqref{MFCGLEmod2nd} with identical initial phases and $B=Z$ value.
Two values of the coupling are selected $\kappa=0.2$ and $0.5$
in Figs.~\ref{QPStransient}(a,b) and \ref{QPStransient}(c,d), respectively.
The heteroclinic connection with the saddle QPS is captured by the
extended model \eqref{MFCGLEmod2nd}, in contrast to \eqref{MFCGLEmod}, which only reproduces
the exponential instability of UIS. For both, the MF-CGLE \eqref{MFCGLE} and Eq.~\eqref{MFCGLEmod2nd},
the final state is a NUIS.
Unsurprisingly, the extended QPR \eqref{MFCGLEmod2nd}
is more accurate for $\kappa=0.2$ than for $\kappa=0.5$, since
we assumed a small $\kappa$ in its derivation.

For $\kappa$ values larger than those in Fig.~\ref{QPStransient}
there is not a saddle QPS but, instead, a
stable QPS branching off from UIS \cite{NK95,CP19}.
Remarkably, this also occurs for the extended model \eqref{MFCGLEmod2nd} at large enough $\kappa$ (not shown).

\section{($N+2M$)-quasi phase reduction: higher-order harmonics}
\label{sechighharm}

In this section we analyze QPR when we let the oscillators
interact through higher-order harmonics:
	\begin{equation}\label{gentheeqM}
	\dot{A_j}=f(A_j)+\kappa \, g(\mathcal{A}^M),
	\end{equation}
where $\mathcal{A}^M$ are mean fields belonging to the set:
\begin{equation}\label{cam}
\mathcal{C}_A^M=\left\{\left\{\overline{|A|^nA^m}\right\}
\cup \left\{\overline{|A|^nA^{*m}}\right\}\right\}_{\substack{n\in\mathbb{Z} \qquad \\ m=1,\ldots,M}}.
\end{equation}
These mean fields are the first $M$ harmonics in $\phi$. We show next that,
provided the subset of \eqref{cam} with $m=M$ is not empty, 
the QPR of Eq.~\eqref{gentheeqM} possesses $N+2M$ degrees of freedom.
In other words, the largest harmonic of $\phi$ in the coupling determines
the number of degrees of freedom of QPR.

We proceed as in the case $M=1$ seeking to close Eq.~\eqref{protophasered}.
For $M>1$ we need to introduce new mean fields 
$B_m=\overline{A^m}$, with $m=1,\ldots,M$ ($B_1\equiv B$). Assuming the $\lambda-\omega$
oscillators are in the neighborhood of the limit cycle at $r=1$, we get:
\begin{eqnarray}
B_m&=&\overline{A^m}=\overline{r^m
e^{im\phi}}\nonumber\simeq\overline{(1+m\delta r)
e^{im(\theta+i\chi_0\delta
r)}} \\
&\simeq& Z_m+m(1+i\chi_0)\overline{\delta r e^{im\theta}}
\end{eqnarray}
where $Z_m=\overline{e^{im\theta}}$ is the $m$-th Kuramoto-Daido order
parameter.
We can express $\overline{\delta r e^{im\theta}}$ in terms of $B_m$ and
$Z_m$:
\begin{equation}
\overline{\delta r e^{im\theta}}\simeq\frac{B_m-Z_m}{m(1+i\chi_0)}
\end{equation}
Applying this equality to the averages $\overline{|A|^nA^m}$ with arbitrary $n$ value
yields: 
\begin{eqnarray}\label{AnAmbar}
\overline{|A|^nA^m}&=&\overline{r^{n+m}e^{im\phi}}\simeq
Z_m+(n+m+im\chi_0)\overline{\delta r e^{im\theta}}\nonumber\\
&\simeq& B_m+\frac{n}{m(1+i\chi_0)}(B_m-Z_m)
\end{eqnarray}
This { relationship} permits to approximate $g(\mathcal{A}^M)$ in Eq.~\eqref{protophasered} in terms of 
the $M$-dimensional complex vectors
$\vec Z=(Z_1,Z_2,\ldots,Z_M)$ and $\vec B=(B_1,B_2,\ldots,B_M)$:
     \begin{equation}\label{gGM}
     g(\mathcal{A}^M)\simeq\Gamma\left(\vec Z,\vec B\right)
     \end{equation}
The evolution of the phases is, therefore, linked to the set of complex mean fields $\{B_m\}_{m=1,\ldots,M}$,
whose evolution equations remain to be determined.
Recalling \eqref{gentheeqM}, we get:
     \begin{equation}
\dot{B}_m
=m\left[\overline{A^{
     m-1}f(A)} +\kappa B_{m-1}g(\mathcal{A}^M)\right] \label{ecuacion}
     \end{equation}
We see that every $B_m$ is 
influenced
by $B_{m-1}$ and by $B_M$  
(and possibly other $B_m$'s) through $g$, see Eq.~\eqref{gGM}.
The first term in the right-hand side of Eq.~\eqref{ecuacion}
is approximated resorting to Eqs.~\eqref{fexpan}, \eqref{nfn} and
\eqref{AnAmbar}. The result depends only on $B_m$ and $Z_m$:
     \begin{equation}
     \overline{A^{m-1}f(A)}=i\Omega B_m+\frac\Lambda m(B_m-Z_m) \nonumber
     \end{equation}
     
Hence, the ($N+2M$)-QPR of Eq.~\eqref{gentheeqM}
is the ($N+2M$)-dimensional set of ordinary differential equations:      
     \begin{subequations}
     \begin{eqnarray}
     \dot{\theta}_j&=&
\Omega+\kappa\operatorname{Im}\left[(1-i\chi_0)\Gamma(\vec Z,\vec B)e^{-i\theta_j}
\right] , \\
     \dot{B}_m&=&\Lambda(B_m-Z_m)+m\left[i\Omega B_m+ \kappa\, \Gamma(\vec Z,\vec B) B_{m-1}\right]\nonumber\\ 
     \end{eqnarray}
\end{subequations}
where $j=1,\ldots,N$, $m=1,\ldots,M$ and $B_0=1$.

As a final note, we mention that it is also possible to deal with a coupling
function $g\left(\overline{|A|}\right)$,
where $g$ is any function \footnote{In this case QPR
can be accomplished using  $\overline{|A|}$ and $\overline{A}$
as the collective variables ($N+3$ degrees of freedom in total).}.

\section{Conclusions}

Phase reduction is a powerful technique that has 
deeply shaped our knowledge on the dynamics
of oscillator ensembles. In spite of its enormous success, the description 
enabled
by reduced phase models breaks down if the coupling between the oscillators is not weak. 
Recently,
some works have extended standard phase reduction 
\cite{matheny19,leon19,wilson_ermentrout_prl19} 
in a perturbative fashion in the coupling constant. These approaches are however condemned to fail
for strong coupling. 

In this work we have given a new twist to the concept of phase reduction,
introducing QPR for all-to-all strongly coupled $\lambda-\omega$ oscillators.
This new reduction procedure exploits the smallness of the collective oscillations
near incoherent states, independently of the coupling strength.
The reduced model has $N+2M$ degrees of freedom corresponding to $M$
dynamical complex variables mediating the interactions
of $N$ phase oscillators, akin to dynamical quorum-sensing models. 
We have studied in detail the case $M=1$, corresponding to interactions
via the first harmonic of the angle. Explicit stability boundaries for
uniform and nonuniform incoherent states have been obtained for ensembles
of Stuart-Landau oscillators. 

Finally, an extension of QPR beyond the lowest order has been obtained for weak coupling.   
Nonetheless, a genuine well-controlled expansion to the next order
remains to be developed. In parallel with this, some sort of generalization
from global
to more complex coupling topologies or more general oscillators appears to be possible as well. 
The case of heterogeneous oscillators ---where traditional phase reduction works perfectly 
(for small couplings)--- is amenable to analysis through QPR and will be the aim of future work.
All in all, we deem QPR as a promising path towards a
comprehensive theory of collective phenomena in oscillator ensembles.

 \begin{acknowledgments}
We thank Juan M.~L\'opez for proofreading our manuscript.
 We acknowledge support by
Agencia Estatal de Investigaci\'on (Spain) and FEDER (EU)
under project No.~FIS2016-74957-P. 
IL acknowledges support by Universidad de Cantabria and 
Government of Cantabria
under the Concepci\'on Arenal programme.
 \end{acknowledgments}


\section*{Appendix: NUIS stability boundary for nonlinear coupling}
\begin{subequations}
	The NUIS stability boundary of \eqref{NLSL} is  determined using the 
Routh-Hurwitz criterion and is given by:
\begin{multline}
	4 (\gamma_n ^2+1) (c_2^2+1) \kappa  (Q^2-1) (\kappa  (n+2)-4)^2+\\
	(4-\kappa  (n+2)) (-4 \gamma_n  c_2+(\gamma_n ^2+1) \kappa  (n 
Q^2+n+2)-4) \\
	(-8 \kappa  (\gamma_n  c_2+n+3)+(\gamma_n ^2+1) \kappa ^2 (-(n^2 
(Q^2-1)-4 n-4))+16)\\
	-4 \kappa  (4 \gamma_n  c_2-(\gamma_n^2+1) \kappa  (n Q^2+n+2)+4)^2>0
	\end{multline}
	\begin{equation}
	\kappa<\frac{4}{2+n}
	\end{equation}
	\begin{equation}
	(\gamma_n^2+1) \kappa ^2 ((n^2 (1-Q^2)+4 n+4))+16-8 \kappa  (\gamma_n  
c_2+n+3)>0
	\end{equation}
	\begin{equation}
	\kappa  ((\gamma_n ^2+1) \kappa (n Q^2+n+2)-4 \gamma_n  c_2-4)>0
	\end{equation}
\end{subequations}

\end{document}